\documentclass[12pt]{article}%
\usepackage{amsmath,latexsym}
\usepackage{graphicx}
\usepackage{amsmath}
\usepackage{amsfonts}
\usepackage{amssymb}%
\begin{document}

\title{{Wormholes in $f(R)$ gravity with a
noncommutative-geometry background}}
   \author{
Peter K F Kuhfittig*\\  \footnote{kuhfitti@msoe.edu}
 \small Department of Mathematics, Milwaukee School of
Engineering,\\
\small Milwaukee, Wisconsin 53202-3109, USA}

\date{}
 \maketitle

\begin{abstract}\noindent
This paper discusses the possible existence
of traversable wormholes in $f(R)$ modified
gravity while assuming a noncommutative-geometry
background, as well as zero tidal forces.  The
first part of the paper aims for an overview via
several shape functions by determining the
corresponding wormhole solutions and their
properties.  The solutions are made complete by
deriving the modified-gravity functions $F(r)$
and $f(R)$, where $F=df/dR$.  It is subsequently
shown that the violation of the null energy
condition can be attributed to the combined
effects of $f(R)$ gravity and noncommutative
geometry.  The second part of the paper
reverses the strategy by starting with a
special form of $f(R)$ and determining the
wormhole solution and the concomitant $F(r)$.
The approach in this paper differs in
significant ways from that of Jamil et al.,
\emph{J. Korean Physical Soc.} \textbf{65}
917 (2014).   \\
\end{abstract}


\textbf{PAC Nos.:} 04.20.Jb; 04.50.Kd
\section{Introduction}\label{E:introduction}

Traversable wormholes, first proposed by Morris
and Thorne \cite{MT88}, are perfectly valid
solutions of the Einstein field equations.  They
have remained somewhat controversial, however,
because their existence depends on a violation
of the null energy condition, requiring the use
of ``exotic matter."  This requirement reduces
the probability of the existence of naturally
occurring wormholes if we adhere to classical
general relativity.  The most obvious
alternative would therefore be a modification
or extension of Einstein's theory.

The best-known candidate for a modification
is $f(R)$ gravity, strongly motivated by its
success in the analysis of large-scale
structures such as galaxies and clusters of
galaxies to account for flat rotation curves
without the need for dark matter and on a
cosmological scale to explain the accelerated
expansion.  Other key motivations are the
possibility of realistic descriptions of strong
gravitational fields near curvature
singularities \cite{fL08} and the possibility of
threading wormholes with ordinary matter
\cite{LO09, pK}, as opposed to the usual exotic
matter.  Another way to help eliminate the need for
exotic matter is to assume a noncommutative-geometry
background.  As discussed below, noncommutative
geometry, an offshoot of string theory, has a
direct physical origin, thereby providing a
simpler motivation.

An interesting counterpart is provided by
$f(R,T)$ gravity \cite{YIa, YIb}.  Here $R$
is the Ricci scalar (Sec. \ref{S:N}) and $T$
is the trace of the energy-momentum tensor.
In Ref. \cite{YIa} the extra curvature
quantities can be interpreted as a
gravitational entity that supports the
wormhole without the need for exotic matter.
Ref. \cite{YIb} considers wormhole geometries
filled with two physically different fluid
configurations, one isotropic and another
anisotropic, also leading to wormhole
solutions without exotic matter.  By contrast,
in the present study, the avoidance of
exotic matter can be attributed to two
geometric rather than physical factors, the
combined effects of $f(R)$ gravity and
noncommutative geometry.  Moreover, the
combined effects may differ from the
individual effects.

In the context of $f(R)$ gravity, another
important variation, discussed in Ref.
\cite{BYA}, is the construction of charged
thin-shell wormholes by surgically grafting
two cylindrically symmetric spacetimes via
the cut-and-paste technique.  Both
logarithmic and exponential forms of $f(R)$
models are used in the stability analysis.
It was concluded that stable solutions are
possible due to the extra curvature
invariants.

It should be stressed that the gravitational
theories in the present paper, $f(R)$
modified gravity and noncommutative geometry
are essentially independent and their effects
on wormhole physics  have been studied
separately.  On the other hand, the theories
are not mutually exclusive and can therefore
be valid simultaneously.  So it is just as
important to study the combined effects as
the effects considered separately.  A brief
overview of these theories and their
incorporation in the present study is
discussed next

\section{Noncommutative geometry and $f(R)$
    modified gravity}\label{S:N}

As noted in the Introduction, this paper
discusses the possible existence of traversable
wormholes by assuming two simultaneous
modifications of gravity, noncommutative geometry
and $f(R)$ modified gravity.  The former refers
to an important outcome of string theory, namely
the realization that coordinates may become
noncommutative operators on a $D$-brane
\cite{eW96, SW99}.  As discussed in Refs.
\cite{NSS06, SS03, mR11}, we are now dealing
with a fundamental discretization of spacetime
due to the commutator $[\textbf{x}^{\mu},
\textbf{x}^{\nu}]=i\theta^{\mu\nu}$, where
$\theta^{\mu\nu}$ is an antisymmetric matrix.
Thus noncommutativity  replaces point-like
objects by smeared objects with the aim of
eliminating the divergences that normally
appear in general relativity.

To describe the smearing, a model that naturally
presents itself is a Gaussian distribution of
minimal length $\sqrt{\alpha}$ instead of the
Dirac delta function \cite{NSS06, mR11, RKRI12,
pK13}.  The more convenient approach used in
this paper is to assume that the energy density
of the static and spherically symmetric and
particle-like gravitational source has the form
\begin{equation}\label{E:rho}
  \rho(r)=\frac{M\sqrt{\alpha}}
     {\pi^2(r^2+\alpha)^2}.
\end{equation}
(See Refs. \cite{LL12} and \cite{NM08} for
further details.)  Here the mass $M$ is diffused
throughout the region of linear dimension
$\sqrt{\alpha}$ due to the uncertainty.
The noncommutative geometry is an intrinsic
property of spacetime and does not depend
on particular features such as curvature.

The other modification, $f(R)$ modified
gravity, replaces the Ricci scalar $R$ in the
Einstein-Hilbert action
\begin{equation*}
  S_{\text{EH}}=\int\sqrt{-g}\,R\,d^4x
\end{equation*}
by a nonlinear function $f(R)$:
\begin{equation*}
   S_{f(R)}=\int\sqrt{-g}\,f(R)\,d^4x.
\end{equation*}
(For a review, see Refs. \cite{fL08, SF10, NO07}.)
Wormhole geometries in $f(R)$ modified gravitational
theories are discussed in Ref. \cite{LO09}.

To describe a spherically symmetric wormhole spacetime,
we take the metric to be \cite{MT88}
\begin{equation}\label{E:line1}
    ds^2=-e^{\Phi(r)}dt^2+\frac{dr^2}{1-b(r)/r}
    +r^2(d\theta^2+\text{sin}^2\theta\,d\phi^2).
\end{equation}
(We are using units in which $c=G=1$.)  Here we
recall that $b=b(r)$ is called the
\emph{shape function} and $\Phi=\Phi(r)$ the
\emph{redshift function}.  For the shape function
we must have $b(r_0)=r_0$, where $r=r_0$ is the
radius of the \emph{throat} of the wormhole.  To
form a wormhole, $b(r)$ must be an increasing
function that also satisfies the \emph{flare-out
condition} $b'(r_0)<1$ \cite{MT88}, as well as
$b(r)<r$ near the throat.  These restrictions
automatically result in the violation of the null
energy condition (and hence the need for ``exotic
matter") in classical general relativity, but not
in $f(R)$ modified gravity, as we will see.

Regarding the redshift function, we normally require
that $\Phi(r)$ remain finite to prevent an event
horizon.  In the present study involving $f(R)$
gravity, we need to assume that $\Phi(r)\equiv
\text{constant}$, so that $\Phi'\equiv 0$.
Otherwise, according to Lobo \cite{LO09}, the
analysis becomes intractable.  On the positive
side, the condition $\Phi'\equiv 0$, called the
\emph{zero-tidal-force} solution in Ref. \cite
{MT88}, is a highly desirable feature for a
traversable wormhole.

Next, let us state the gravitational field
equations in the form used by Lobo and
Oliveira \cite{LO09}:
\begin{equation}\label{E:FE1}
   \rho(r)=F(r)\frac{b'(r)}{r^2},
\end{equation}
\begin{equation}\label{E:FE2}
  p_r(r)=-F(r)\frac{b(r)}{r^3}+F'(r)
  \frac{rb'(r)-b(r)}{2r^2}-F''(r)
     \left[1-\frac{b(r)}{r}\right].
\end{equation}
and
\begin{equation}\label{E:FE3}
  p_t(r)=-\frac{F'(r)}{r}\left[1-\frac{b(r)}{r}
  \right]+\frac{F(r)}{2r^3}[b(r)-rb'(r)],
\end{equation}
where $F=\frac{df}{dR}$.  The curvature scalar $R$ is
given by
\begin{equation}\label{Ricci}
   R(r)=\frac{2b'(r)}{r^2}.
\end{equation}

Some aspects of wormholes in $f(R)$ modified
gravity with a noncommutative-geometry background
have already been discussed in Ref. \cite{JRMKAM}
based on the Gaussian distribution.  The emphasis
is on wormhole solutions obtained from the
power-law form $f(R)=aR^n$, particularly for higher
powers.  This paper uses a different approach: we
assume various forms of the shape function,
together with a noncommutative-geometry background
via Eq. (\ref{E:rho}) to determine not only the
wormhole solutions and their surprisingly similar
properties, but also the functions $F(r)$ and
$f(R)$ needed for a complete solution.  It is
shown in the process that the energy violations
can be attributed to the combined effects of
noncommutative geometry and $f(R)$ gravity,
without requiring exotic matter.  These ideas
are discussed in Sec. \ref{S:specific}.
Sec. \ref{S:f(R)} reverses the strategy by
starting with a form of $f(R)$ and determining
both the wormhole solution and the function
$F(r)$.

\section{Solutions based on specific shape
functions}\label{S:specific}

Our starting point is the fairly standard shape
function
\begin{equation}\label{E:shape1}
   b(r)=r_0\left(\frac{r}{r_0}\right)^{\beta},
   \quad0<\beta <1,
\end{equation}
which includes the important special case
$\beta =\frac{1}{2}$:
\begin{equation}\label{E:shape2}
   b(r)=\sqrt{r_0r}.
\end{equation}
In both cases, $\text{lim}_{r\rightarrow \infty}
b(r)/r=0$, so that the spacetimes are
asymptotically flat.  Also, in both cases,
$b(r_0)=r_0$.  Since
\begin{equation}\label{E:bprime}
   b'(r)=\beta\left(\frac{r}{r_0}\right)
   ^{\beta -1},
\end{equation}
we have $b'(r_0)=\beta <1$, so that the
flare-out condition is satisfied.  It also
follows from Eqs. (\ref{E:rho}) and
(\ref{E:FE1}) that
\begin{equation*}
   \frac{M\sqrt{\alpha}}{\pi^2(r^2+\alpha)^2}
   =F(r)\frac{\beta(r/r_0)^{\beta-1}}{r^2}
\end{equation*}
and
\begin{equation}\label{E:F(r)1}
   F(r)= \frac{M\sqrt{\alpha}}{\pi^2(r^2
   +\alpha)^2}\frac{r^{3-\beta}}
   {\beta r_0^{1-\beta}}.
\end{equation}
The expression for $F(r)$ allows us to compute
both the radial and transverse pressures from
Eqs. (\ref{E:FE2}) and (\ref{E:FE3}),
respectively.  We also observe that in making
use of Eq. (\ref{E:rho}), the right-hand sides
of Eqs. (\ref{E:FE1})-(\ref{E:FE3}) remain
intact.  Only the stress-energy tensor is
modified; so the length scales are macroscopis
\cite{NSS06}.

As noted earlier, for any wormhole, the
flare-out condition $b'(r_0)<1$ must be met.
For a Morris-Thorne wormhole this condition
is sufficient to yield a violation of the
null energy condition (NEC), $\rho +p_r<0$,
at the throat: if $F(r)\equiv 1$ in Eqs.
(\ref{E:FE1}) and (\ref{E:FE2}), then
\begin{equation}\label{E:NEC1}
   \rho(r_0)+p_r(r_0)=\frac{r_0b'(r_0)
   -b(r_0)}{r_0^3}<0
\end{equation}
since $b(r_0)=r_0$.  For $f(R)$ gravity, this
conclusion no longer holds because of the
dependence on $F(r)$.  The violation must
therefore be checked separately.

Since the general expression for $\rho+p_r$
is rather lengthy, we will check the violation
only at the throat and, by continuity, in its
immediate vicinity:
\begin{multline}\label{E:NEC2}
   \rho(r_0)+p_r(r_0)=
   \frac{M\sqrt{\alpha}}{\pi^2(r_0^2+\alpha)^2}
   -\frac{M\sqrt{\alpha}}{\pi^2(r_0^2+\alpha)^2}
   \frac{r_0^{3-\beta}}{\beta r_0^{1-\beta}}
   \frac{b(r_0)}{r_0^3}\\+F'(r_0)
   \frac{r_0b'(r_0)-b(r_0)}{2r_0^2}
   -F''(r_0)\left(1-\frac{b(r_0)}{r_0}\right)\\
   =\frac{M\sqrt{\alpha}}{\pi^2(r_0^2+\alpha)^2}
   \left(1-\frac{1}{\beta}\right)+F'(r_0)
   \frac{\beta-1}{2r_0}
\end{multline}
since $b(r_0)=r_0$ and $b'(r_0)=\beta$.
Substituting
\begin{equation*}
   F'(r_0)=\frac{-r_0^{2-\beta}[(\beta+1)r_0^2
   +\alpha(\beta-3)]}{(r_0^2+\alpha)^2},
\end{equation*}
Eq. (\ref{E:NEC2}) can be reduced further to become
\begin{equation}\label{E:NEC3}
   \rho(r_0)+p_r(r_0)=
   \frac{M\sqrt{\alpha}}{\pi^2(r_0^2+\alpha)^2}
   \left(1-\frac{1}{\beta}+\frac{1}{2\beta}
   \frac{r_0^2-\beta^2r_0^2+4\alpha\beta-3\alpha
   +\alpha\beta^2}{r_0^2+\alpha}\right).
\end{equation}
The final form is
\begin{equation}\label{E:NEC4}
   \rho(r_0)+p_r(r_0)=
   \frac{M\sqrt{\alpha}}{\pi^2(r_0^2+\alpha)^2}
   \frac{(\beta-1)[(1-\beta)r_0^2+\alpha(5-\beta)]}
   {2\beta(r_0^2+\alpha)}<0.
\end{equation}
since $\beta<1$.  So the NEC has been violated.
(The dependence on Eqs. (\ref{E:rho}) and
(\ref{E:FE1}) shows that this conclusion can be
attributed to the combined effects of
noncommutative geometry and $f(R)$ gravity,
without requiring exotic matter.)

In the special case $\beta=\frac{1}{2}$, which
will be needed later, we get
\begin{equation}\label{E:NEC5}
  \rho(r_0)+p_r(r_0)=
   \frac{M\sqrt{\alpha}}{\pi^2(r_0^2+\alpha)^2}
   \left(-1+\frac{3r_0^2-5\alpha}
   {4r_0^2+4\alpha}\right)<0.
\end{equation}
So all the conditions for a traversable wormhole
have been met.

To finish the discussion, we need to return to
Eq. (\ref{E:F(r)1}),
\begin{equation}\label{E:F(r)2}
 F(r)=\frac{M\sqrt{\alpha}}{\pi^2(r^2+\alpha)^2}
 \frac{1}{\beta r^{\beta-3}r_0^{1-\beta}}
\end{equation}
and
\begin{equation}\label{E:R}
   R(r)=\frac{2b'}{r^2}=2\beta r^{\beta-3}
   r_0^{1-\beta}.
\end{equation}
So making use of Eq. (\ref{E:F(r)2}), $F(r)$ can
now be written
\begin{equation}\label{E:F(r)3}
   F(r)=\frac{2}{2\beta r^{\beta-3}
   r_0^{1-\beta}}\rho=\frac{2}{R}\rho.
\end{equation}

To derive $F(R)$, we need to invert $R(r)$ to
obtain $r(R)$.  From Eq. (\ref{E:R}),
\begin{equation}\label{E"r(R)1}
   r(R)=\left(\frac{2\beta r_0^{1-\beta}}
   {R}\right)^{1/(3-\beta)}.
\end{equation}
Also needed is $r_0(R)$:
\begin{equation}\label{E:Rzero1}
   R_0=R(r_0)=\frac{2\beta}{r_0^2},
\end{equation}
so that
\begin{equation}\label{E:rzero}
   r_0=\left(\frac{2\beta}{R_0}\right)^{1/2}.
\end{equation}
Thus
\begin{equation}\label{E:radius1}
   r=\left(\frac{2\beta(2\beta/R_0)^
   {\frac{1}{2}(1-\beta)}}{R}\right)
   ^{1/(3-\beta)}.
\end{equation}
Eq. (\ref{E:F(r)3}) now yields
\begin{equation}\label{E:F(R)1}
  F(R)=\frac{M\sqrt{\alpha}}{\pi^2}
  \frac{2}{R}\left (\left[\frac{R}{2\beta}
  \left(\frac{2\beta}{R_0}\right)^
  {\frac{1}{2}(\beta-1)}\right]
  ^{2/(\beta-3)}+\alpha\right )^{-2}.
\end{equation}
For the special case $\beta =\frac{1}{2}$,
we get
\begin{equation}\label{E:F(R)2}
   F(R)=\frac{2M\sqrt{\alpha}}{\pi^2}
   \frac{1}{R}\left[(RR_0^{1/4})^{-4/5}
   +\alpha\right ]^{-2}.
\end{equation}
Recalling that $\frac{df}{dR}=F$, integration
yields the following closed form:
\begin{equation}\label{E:f(R)1}
   f(R)=\frac{10M\sqrt{\alpha}}{\pi^2}
   \frac{(\alpha R_0^{1/5}R^{4/5}+1)\,
   \text{ln}\,(\alpha R_0^{1/5}R^{4/5}+1)
   -\alpha R_0^{1/5}R^{4/5}}
   {2\alpha^2(\alpha R_0^{1/5}R^{4/5}+1)}+C.
\end{equation}
The graph of $f(R)$ is shown in Fig. 1 for $C=0$.
\begin{figure}[tbp]
\begin{center}
\includegraphics[width=0.8\textwidth]{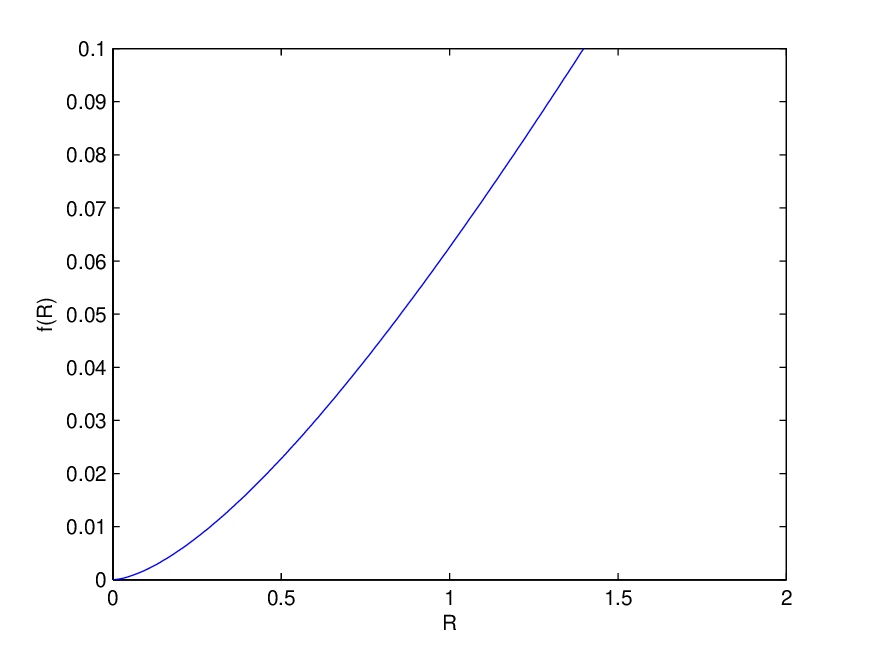}
\end{center}
\caption{The graph of $f(R)$ corresponding to
the shape function $b(r)=\sqrt{r_0r}$.}
\end{figure}

For the purpose of comparison and gaining an overview,
we now consider another shape function,
\begin{equation}\label{E:shape3}
   b(r)=r_0+ar_0\left(1-\frac{r_0}{r}\right),
   \quad 0<a<1.
\end{equation}
This time we get
\begin{equation}\label{ED:bprime2}
   b'(r)=a\left(\frac{r_0}{r}\right)^2,
\end{equation}
so that $b'(r_0)=a<1$, and the flare-out condition
has been met.  Next, we find that
\begin{equation}\label{E:F(r)4}
   F(r)=\frac{M\sqrt{\alpha}}{a\pi^2r_0^2}
   \frac{r^4}{(r^2+\alpha)^2},
\end{equation}
while
\begin{equation}\label{E:NEC6}
   \left. \rho(r)+p_r(r)\right|_{r=r_0}=
   \frac{M\sqrt{\alpha}}{\pi^2(r_0^2+\alpha)^2}
   \left[\frac{a-1}{a}\left(1+
   \frac{2\alpha}{r_0^2+\alpha}\right)\right]<0
\end{equation}
since $a<1$.  So the NEC is violated at the throat.

To obtain $f(R)$, we proceed as before, starting
with $F(r)$:
\begin{equation}\label{E:F(r)5}
   F(r)=\frac{r^2\rho}{b'(r)}=
   \frac{r^4}{ar_0^2}\rho,
\end{equation}
\begin{equation}\label{E:R2}
   R(r)=\frac{2b'}{r^2}=\frac{2ar_0^2}{r^4},
\end{equation}
and
\begin{equation}\label{E"Rzero2}
   R_0=R(r_0)=\frac{2a}{r_0^2},
\end{equation}
resulting in $r_0^2=2a/R_0$.  To obtain $r(R)$,
we note that $r^4R=2ar_0^2=2a(2a/R_0)$ and
\begin{equation}\label{E:rsquared}
   r^2=\frac{2a}{\sqrt{R_0R}}.
\end{equation}
Returning to $F(r)$,
\begin{equation}\label{E:F(r)5}
   F(r)=\frac{r^4}{a(2a/R_0)}\rho=
   \frac{R_0}{2a^2}(r^2)^2\rho
\end{equation}
and
\begin{equation}\label{E:F(R)3}
   F(r)=\frac{R_0}{2a^2}\frac{4a^2}{R_0R}
   \rho=\frac{2}{R}\rho,
\end{equation}
in agreement with Eq. (\ref{E:F(r)3}).  So
\begin{equation}\label{E:F(R)4}
   F(R)=\frac{2}{R}\frac{M\sqrt{\alpha}}
   {\pi^2}\left(\frac{2a}{\sqrt{R_0R}}
   +\alpha\right)^{-2}.
\end{equation}
This yields the following closed form for
$f(R)$:
\begin{equation}\label{E:f(R)2}
   f(R)=\frac{4M\sqrt{\alpha}}{\pi^2}
   \frac{(\alpha\sqrt{R_0R}+2a)\,\text{ln}\,
   (\alpha\sqrt{R_0R}+2a)-\alpha\sqrt{R_0R}}
   {\alpha^2(\alpha\sqrt{R_0R}+2a)}+C.
\end{equation}

Continuing the comparison, suppose we consider the
shape function
\begin{equation}\label{E:shape4}
   b(r)=r_0+a(r-r_0),\quad a<1,
\end{equation}
leading to $b'(r)=a<1$, so that the flare-out
condition is satisfied.  This time let us simply
list the basic features:
\begin{equation}
   F(r)=\frac{M\sqrt{\alpha}}{\pi^2a}
      \frac{r^2}{(r^2+\alpha)^2},
\end{equation}
\begin{equation}
   F'(r)=\frac{M\sqrt{\alpha}}{\pi^2a}
      \frac{-2r(r^2-\alpha)}{(r^2+\alpha)^3},
\end{equation}
and
\begin{equation}\label{E:NEC7}
  \rho(r_0)+p_r(r_0)=\frac{M\sqrt{\alpha}}
  {\pi^2(r_0^2+\alpha)^2}\frac{a-1}{a}\left(
  1-\frac{r_0^2-\alpha}{r_0^2+\alpha}\right).
\end{equation}
So $\rho(r_0)+p_r(r_0)<0$, since $a<1$.
Finally, $F(r)=(2/R)\rho$, as before,
leading to
\begin{equation}
  F(R)=\frac{2M\sqrt{\alpha}}{\pi^2}
     \frac{1}{R(2a/R+\alpha)^2}
\end{equation}
and
\begin{equation}
  f(R)=\frac{2M\sqrt{\alpha}}{\pi^2}
  \frac{(\alpha R+2a)\,\text{ln}\,
  (\alpha R+2a)-\alpha R}
  {\alpha^2(\alpha R+2a)}+C.
\end{equation}

These comparisons have shown that all the
shape functions have yielded surprisingly
similar results.  The reason is that the
violation of the NEC at the throat will
occur under fairly general conditions
since $b(r_0)=r_0$ and $b'(r_0)<1$:
\begin{multline}
   \left .\rho(r)+p_r(r)\right |_{r=r_0}=
   \frac{M\sqrt{\alpha}}{\pi^2}\frac{1}
   {(r^2+\alpha)^2}-F(r)\frac{b}{r^3}\\
   +\frac{F'(r)}{2r^2}(b'r-b)
   -F''(r)\left .\left(1-\frac{b}{r}
   \right)\right|_{r=r_0}
\end{multline}
will be less than zero if $F(r_0)>0$ and
$F'(r_0)>0$, primarily because $\alpha
\ll 1$.

An interesting contrast is provided by the shape
function
\begin{equation}
   b(r)=\frac{r_0^2}{r}.
\end{equation}
Here we obtain
\begin{equation}
   F(r)=-\frac{M\sqrt{\alpha}}{\pi^2r_0^2}
      \frac{r^4}{(r^2+\alpha)^2}
\end{equation}
and
\begin{equation}
   F'(r)=-\frac{M\sqrt{\alpha}}{\pi^2r_0^2}
     \frac{4\alpha r^3}{(r^2+\alpha)^3}.
\end{equation}
The result is
\begin{equation}\label{E:contrast1}
   \rho(r_0)+p_r(r_0)=\frac{2M\sqrt{\alpha}}
   {\pi^2(r_0^2+\alpha)^2}\left(1+
   \frac{2\alpha}{r_0^2+\alpha}\right)>0.
\end{equation}
Since the NEC is met, we do not get a wormhole.

On the other hand, Einstein's theory with a
noncommutative-geometry background yields
\begin{equation}\label{E:contrast2}
    \left .\rho(r_0)+p_r(r_0)=
    \frac{M\sqrt{\alpha}}{\pi^2(r^2+\alpha)^2}
    -\frac{r_0^2/r}{r^3}\right|_{r=r_0}<0
\end{equation}
since $\alpha \ll 1$.  Here the NEC is indeed
violated.

As noted by Lobo and Oliveira \cite{LO09}, if
we assume that the matter threading the wormhole
satisfies the NEC, it is the higher-order
curvature terms that sustain the wormhole.  It
is also clear from Eqs. (\ref{E:NEC4}),
(\ref{E:NEC5}), (\ref{E:NEC6}), and
(\ref{E:NEC7}) that noncommutative geometry
is also a contributing factor due to the small
value of $\alpha$.  We can see from Eqs.
(\ref{E:contrast1}) and ({\ref{E:contrast2}),
however, that the combined effects of $f(R)$
gravity and noncommutative geometry may differ
from the individual effects.

\section{Solutions based on $f(R)$}\label{S:f(R)}

Ref. \cite{JRMKAM} considered the form $f(R)=aR^n$
and obtained a number of wormhole solutions
corresponding to different values of $n$.  Observe
that $\frac{df}{dR}=F(R)=bR^{n-1}$ for some constant
$b$.  The case $n=1$ corresponds to Einstein gravity
(again assuming a noncommutative-geometry background).
In this section we will consider $R^2$ (quadratic)
gravity, in part because this is believed to be a
physically viable model \cite{ABS10}.

So we start with
\begin{equation}
   \frac{df}{dR}=aR=\frac{2ab'}{r^2}=F.
\end{equation}
From Eq. (\ref{E:FE1}),
\begin{equation*}
   b'=\frac{r^2\rho}{F}
   =\frac{r^2\rho}{2ab'/r^2}.
\end{equation*}
Thus
\begin{equation}\label{E:derivative}
  b'(r)=\frac{r^2\sqrt{\rho(r)}}{\sqrt{2a}}
\end{equation}
Continuing with the wormhole solution,
\begin{equation}
   b(r)=\left(\frac{M\sqrt{\alpha}}
   {2a\pi^2}\right)^{1/2}
   \int\frac{r^2dr}{r^2+\alpha}
   =\frac{\sqrt{M}\alpha^{1/4}}
   {\sqrt{2a}\pi}\left(r-\sqrt{\alpha}\,
   \text{tan}^{-1}\frac{r}{\sqrt{\alpha}}
   +C\right).
\end{equation}
To satisfy the condition $b(r_0)=r_0$, we must have
\begin{equation*}
  C=r_0\frac{\sqrt{2a}\pi}{\sqrt{M}\alpha^{1/4}}
  -r_0+\sqrt{\alpha}\,\text{tan}^{-1}
     \frac{r_0}{\sqrt{\alpha}},
\end{equation*}
so that
\begin{equation}
   b(r)=\frac{\sqrt{M}\alpha^{1/4}}
   {\sqrt{2a}\pi}\left(r-\sqrt{\alpha}\,
   \text{tan}^{-1}\frac{r}{\sqrt{\alpha}}+
   r_0\frac{\sqrt{2a}\pi}{\sqrt{M}\alpha^{1/4}}
  -r_0+\sqrt{\alpha}\,\text{tan}^{-1}
     \frac{r_0}{\sqrt{\alpha}}
   \right).
\end{equation}
Next,
\begin{equation}
   b'(r)=\frac{\sqrt{M}\alpha^{1/4}}
   {\sqrt{2a}\pi}
   \left(1-\frac{1}{1+r^2/\alpha}\right).
\end{equation}
So $b'(r)>0$ and $b'(r)<1$ since
$\alpha\ll 1$.  The flare-out condition is
thereby satisfied.

It remains to determine $F(r)$.  Since $b'=
r^2\rho(r)/F$, we have from Eq.
(\ref{E:derivative}),
\begin{equation*}
  \frac{r^2\sqrt{\rho(r)}}{\sqrt{2a}}=
  \frac{r^2\rho(r)}{F};
\end{equation*}
hence
\begin{equation}
   F(r)=\sqrt{2a}\sqrt{\rho(r)}=\sqrt{2a}
   \frac{\sqrt{M}\alpha^{1/4}}{\pi}
   \frac{1}{r^2+\alpha}.
\end{equation}
Alternatively,
\begin{equation}
  \frac{df}{dR}=aR=\frac{2ab'}{r^2}=
  \frac{2a}{r^2}\frac{r^2\sqrt{\rho(r)}}{
  \sqrt{2a}}=\sqrt{2a}\sqrt{\rho(r)},
\end{equation}
as before.  So $\frac{df}{dR}$ reduces to
$F(r)$.

\section{Results and Discussion}

This paper discusses the possible existence of
traversable wormholes in the context of $f(R)$
modified gravity, given a
noncommutative-geometry background based on
the Lorentzian distribution in Eq. (\ref{E:rho}).

The first part of this paper provides an
overview by using several specific shape
functions to show that the corresponding
wormhole solutions share some fairly
general common properties.  An original
objective was to obtain complete wormhole
solutions by determining the modified-gravity
functions $F(r)$ and $f(R)$.  The second part
of the paper reverses the strategy by starting
with $R^2$ gravity, i.e., the power-law form
$f(R)=aR^2$, which is considered to be a
physically viable model \cite{ABS10}.  This
is followed by a determination of the wormhole
solution, including  the function $F(r)$.
Zero tidal forces are assumed throughout.

Implicit in the discussion is that these
gravitational theories are essentially
independent, so that their effects can be
studied separately.  On the other hand,
they are not mutually exclusive, so it
becomes equally important to study the
combined effects.  In particular, it is
shown that the energy violation can be
attributed to the combined effects of
$f(R)$ gravity and noncommutative
geometry, thereby avoiding the need
for exotic matter.  Moreover, the combined
effects may differ from the individual
effects.

\section{Conclusions}

Some aspects of wormholes in $f(R)$ modified
gravity with a noncommutative-geometry
background have already been considered in
Ref. \cite{JRMKAM}.  The discussion is based
on a Gaussian distribution and concentrates
mainly on higher powers of the power-law
form $f(R)=aR^n$.  By contrast, the purpose
of the present study is to seek more complete
solutions by determining the modified-gravity
functions $F(r)$ and $f(R)$.

The two gravitational theories are independent
without being mutually exclusive.  As noted
above, the combined effects may differ from
the individual effects, as exemplified by the
discussion of the violation of the null energy
condition.

\end{document}